\documentclass[conference]{IEEEtran} 
\pdfoutput=1

\usepackage{graphicx,times, epsfig,amsmath} 
\usepackage{amssymb,amsthm}		
\usepackage{stfloats}
\usepackage{slashbox}

\newcommand{\vect}[1]{\underline{\mathbf{#1}}}

\IEEEoverridecommandlockouts

\begin{document}

\title{Outage Probability and Capacity for Two-Tier Femtocell Networks by Approximating Ratio of Rayleigh and Log Normal Random Variables}

\author{\IEEEauthorblockN{Sumudu Samarakoon, Nandana Rajatheva, Mehdi Bennis and Matti Latva-aho}
\IEEEauthorblockA{Department of Communications Engineering, University of Oulu, Finland}
\IEEEauthorblockA{\{sumudu, rrajathe, bennis, matti.latva-aho\}@ee.oulu.fi}
\vspace{-10mm}
\thanks{The research has been funded in part by the LOCON Project, TEKES, Finland.}
}

\maketitle
\nopagebreak[4]
\begin{abstract}
This paper presents the derivation for per-tier outage probability of a randomly deployed femtocell network over an existing macrocell network. The channel characteristics of macro user and femto user are addressed by considering different propagation modeling for outdoor and indoor links. Location based outage probability analysis and capacity of the system with outage constraints are used to analyze the system performance. To obtain the simplified expressions, approximations of ratios of Rayleigh random variables (RVs), Rayleigh to log normal RVs and their weighted summations, are derived with the verifications using simulations.
\end{abstract}
\section{Introduction}\label{sec:intro}
The needs for high capacity, data rates and better quality of service in wireless communication grow by each day. Although the overall demand is high, it is not always distributed uniformly over large areas. Smaller regions such as indoor environment are most likely to have a higher concentration of demand compared to large areas. The concept of femtocells - so called home base stations - is a promising solution, by providing closer wireless links between the transmitter and receiver \cite{pap:femtosurvey}. Femtocell is a low powered personal indoor base station (BS) with short coverage distance. It communicates with user equipment (UE) through wireless links while backhauls to operator network are over internet \cite{pap:femtosurvey},\cite{pap:uplinkcapinf}. 
Femtocells are deployed at traffic hot spots as a two-tier network to improve the overall capacity \cite{pap:hotLAN3G}.

In \cite{pap:anamccdma}, a performance evaluation of multi-carrier coded-division multiple access (MC-CDMA) system with two-tier network is carried out for an arbitrarily distributed femtocells in a Rayleigh faded environment.
The derivation of the semianalytical probability density function (PDF) of the downlink signal to interference and noise ratio (SINR) for a femtocell network is presented in \cite{pap:semianapdf}. Uncoordinated deployment of femto base stations (FBS) with log normal shadow fading is used to model the femtocell indoor environment.
The effect of co-channel interference in a two-tier network is investigated in \cite{pap:perfanatwotier}. 
Using the same propagation model for both indoor and outdoor channels, the system performance is analyzed based on the outage probability and capacity for both macro and femtocell user equipment (MUE and FUE).

In this paper, the two-tier network model in \cite{pap:perfanatwotier} is modified to address the behavioral difference of indoor and outdoor propagations.
The indoor propagation has slow variations, thus the links between FBS and UE are modeled as log normal fading channels.
The outdoor propagation with multi-path and non line-of-sight conditions, channel between MBS and UE is modeled with Rayleigh fading.
With this distinction, we address a system which is closer to the real world scenario compared to the work done in \cite{pap:perfanatwotier}.
The per-tier outage probabilities and capacity derivations provided there are no longer valid for our system.
Thus, we analyze the system performance and limitations for co-channel femtocell deployment based on our approach.
In addition, we provide approximations on PDFs and weighted summations, for the ratios of Rayleigh to Rayleigh and Rayleigh to log normal RVs. This mathematical approximation is verified with the simulations, and it can be applied for any other application, which involves Rayleigh and log normal random variables (RVs).

The rest of the paper is organized as follows. Section \ref{sec:sysmodel} describes the system model in detail and section \ref{sec:theory} examines the derivation of per-tier outage probabilities. Section \ref{sec:combined} is related to the approximation of PDFs of ratios of Rayleigh and log Normal random variables (RV) and weighted summations of those. Numerical results of both simulations and analytical approach are provided in section \ref{sec:results}. The conclusions are given in section \ref{sec:conclu}.

\section{System Model}\label{sec:sysmodel}

We consider a system where randomly distributed circular femtocells are overlaid in a hexagonal macrocell. The macrocell base station (MBS) is located at the center of the macrocell with radius $R_m$. The macrocell users (i.e. user equipment - MUE) are uniformly distributed over the macrocell region. The downlink channel between MBS and MUE undergoes exponential path loss ($\alpha$) and Rayleigh fading ($\phi$) \cite{book:fastfading}.

\begin{figure}[!t]
	\centering
	\includegraphics[scale=0.365]{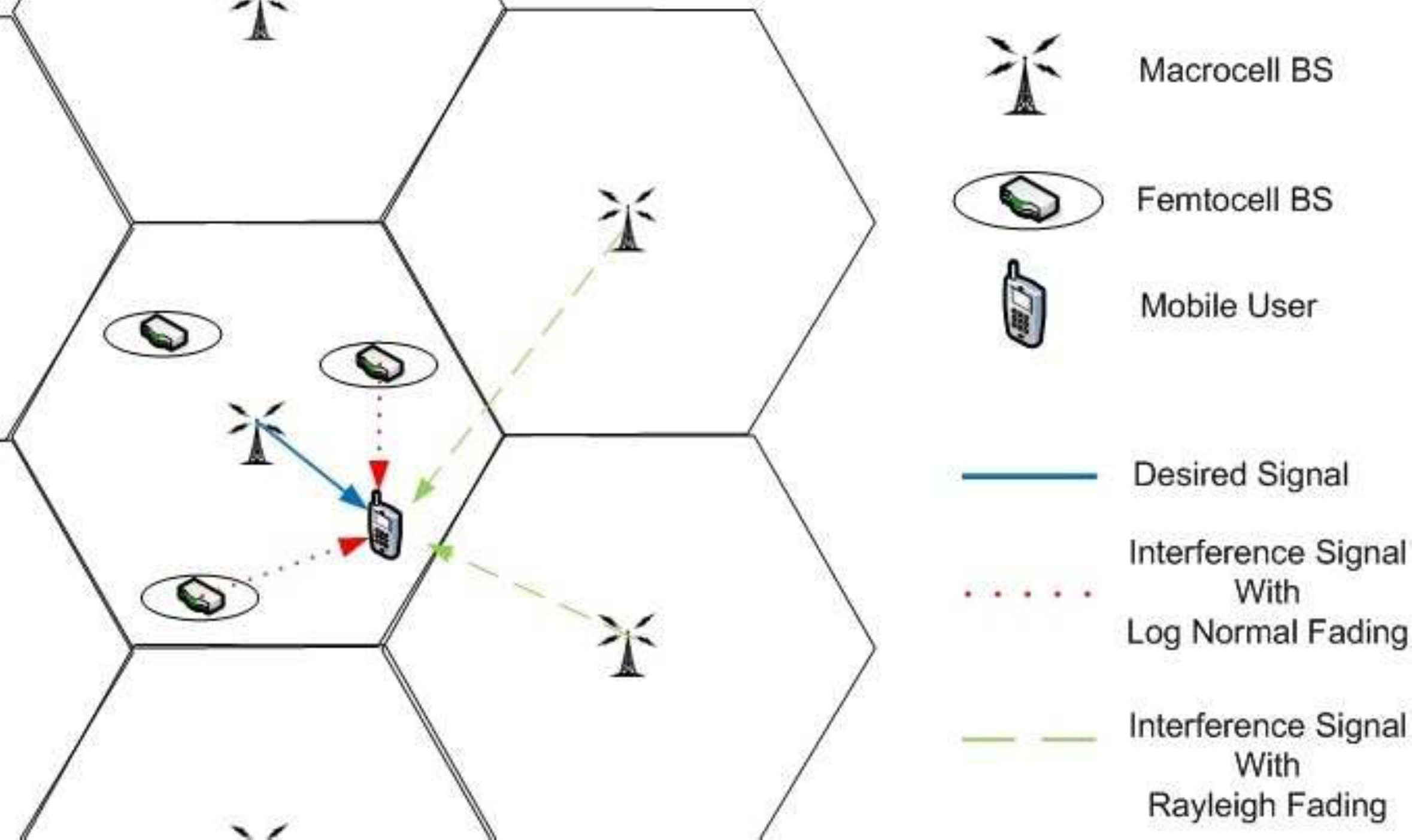}
	\caption{Macrocell - Femtocell Layout}
	\label{fig:macro_femto_layout}
\end{figure}

Femtocells with radius of $R_f$ are distributed in the macrocell following a spatial Poisson point process (SPPP) with intensity of $\lambda_f$ at a given time \cite{pap:uplinkcapinf}. It is assumed that only FBSs which are in the same macrocell interfere with MUEs or femtocell users (i.e. femto user equipment - FUE). The link between a user and FBS suffers from an exponential path loss ($\beta$), wall penetration loss ($W$) and log normal fading ($\psi$) \cite{book:lognormalfading},\cite{pap:pathmodel}.  Fig. \ref{fig:macro_femto_layout} illustrates the layout of the system. The femtocells are assumed to operate under closed access.

All the indoor and outdoor channel fading are assumed to be independent from the positions of BSs and UEs.
The effects of the thermal noise are neglected due to the involvement of a higher number of BSs
\cite[p.428-433]{book:rrmfwn}. 

In terms of notation we consider the $j$-th MBS is $M_j$ and $i$-th FBS is $F_i$. The MBS of the interested macrocell is $M_0$ and the set of neighboring MBSs is $\Omega=\{M_j|j\neq0\}$. For femtocells, $F_0$ is the FBS which serves the desired FUE and $\Lambda=\{F_i\}$ is the set of all the femtocells in the 0-th MBS. $\mathbb{P}(\cdot)$ function denotes the calculated probability.

\section{Per-tier Outage Probability and Capacity}\label{sec:theory}

The worst case scenario of the interference is considered by assuming that all BSs are using the same frequency band for their communication. 
Therefore, the set of interfering MBSs with 0-th MBS - MUE communication is $\Omega_m=\Omega$ (all the neighboring MBSs) and, for FBS - FUE communication is $\Omega_f=(\Omega\cup M_0)$ (all the MBS).
The interference from MBSs on MUE ($I_{m,m}$) and FUE ($I_f,m$) can be calculated by
\begin{align}\label{eqn:interf_macro}
	&I_{m,m}=\sum_{j{\in}\Omega_m}P_j\psi_jD_j^{-\alpha}~;
	&I_{f,m}=\sum_{j{\in}\Omega_f}WP_j\psi_jD_j^{-\alpha},
\end{align}
where $P_j$ is transmission power of MBS $M_j$, and $D_j$ is the distance between $M_j$ and the UE with outdoor path loss exponent $\alpha$. $W$ is the wall penetration loss. The random variable (RV) $\psi_j$ denotes the fading component of the outdoor link with Rayleigh distribution with parameter $\sigma_j$. 
The interference from FBSs on MUE ($I_{m,f}$) and FUE ($I_{f,f}$) are given by 
\begin{align}\label{eqn:interf_femto}
	&I_{m,f}=\sum_{i{\in}\Lambda_m}WP_i\phi_iD_i^{-\beta}~;
	&I_{f,f}=\sum_{i{\in}\Lambda_f}W^2P_i\phi_iD_i^{-\beta},
\end{align}
where $\Lambda_m=\Lambda$ and $\Lambda_f=(\Lambda-F_0)$. The RV $\phi_i$ represents the fading of indoor links, and is assumed to be $LN(0,\sigma_i^2)$. $\beta$ is the indoor path loss exponent.

\subsection{Macrocell Outage Probability}

The SIR of the  MUE which is located at the position $\vect{r}$ is given by,
\begin{equation}\label{eqn:sir_macro}
	\text{SIR}_{m,\vect{r}}=\frac{P_0\psi_0r^{-\alpha}}{I_{c,c}+I_{c,f}}
\end{equation}
where $r=||\vect{r}||$. The outage probability for MUE 
with co-channel density of $\lambda_f$ is
\begin{equation}\label{eqn:outage_macro}
	q_{m,\vect{r}}(\lambda_f)=\mathbb{P}\{\text{SIR}_{m,\vect{r}}<\gamma_m\}=\mathbb{P}\{Y>\frac{1}{\gamma_m}\}=\overline{F_Y}(y)
\end{equation}
where $\gamma_c$ is the target SIR for a MUE. 
In order to simplify the calculation, we consider the aggregated interference of MUE $(Y)$ rather than the SIR.
$\overline{F_Y}(y)$ is the complementary cumulative distribution function (CCDF) of $Y$.
$\overline{F_Y}(y)$ does not have a closed form solution, but can be approximated by modifying the femtocell distribution. Macrocell region (area is $|H|$) is equally divided into $N$ sub regions and the probability of a femtocell occurrence within a sub region is $p=\frac{\lambda_f|H|}{N}$. Therefore Binomial RVs ${\bf X}$ (${\bf x}_k=[x_{k1},…,x_{kN}]$ and $\mathcal{X}=\{{\bf x}_k\}$) with probability $p$ to be equal to 1 is assigned to each sub region to represent the femtocell configuration \cite{pap:perfanatwotier}. By considering the weighted sum of $Y$ for given configuration ${\bf x}_k$ over $\mathcal{X}$, we can calculate $\overline{F_Y}(y)$ as,
\begin{equation}\label{eqn:ccdf}
	\overline{F_Y}(y)=\sum_{x_k{\in}\mathcal{X}}\overline{F_{Y|X}}(y|x_k)\mathbb{P}(X=x_k)
\end{equation}
The conditional distribution can be expressed as 
\begin{multline}\label{eqn:ccdf_condi}
\overline{F_{Y|X}}(y|x_m)=\\ \mathbb{P}\biggl\{\sum_{j{\in}\Omega_m}\frac{P_j\psi_jD_j^{-\alpha}}{P_0\psi_0r^{-\alpha}}+\sum_{i{\in}\Lambda_m}\frac{WP_i\phi_iD_i^{-\alpha}x_{ki}}{P_0\psi_0r^{-\alpha}}>y\bigg|{\bf \underline{x}}=x_k\biggr\}
\end{multline}
Considering equal subdivision, $\mathbb{P}({\bf X=x}_m)$ is calculated using
\begin{equation}
	\mathbb{P}({\bf X=x}_k)=p^{\sum_{i{\in}\mathcal{N}^{x_{k_i}}}}(1-p)^{N-\sum_{i{\in}\mathcal{N}^{x_{k_i}}}}
\end{equation}
We will show in Section \ref{sec:combined}
, that the ratios $\frac{\psi_j}{\psi_0}$ and $\frac{\phi_i}{\psi_0}$ are approximated as log normal RVs. Thus 
 (\ref{eqn:ccdf_condi}) becomes a weighted summation of log normal RVs, which 
can be simplified into single log normal RV by applying Fenton - Wilkinson's method \cite{pap:fenton},\cite{pap:apprxlognorm}.  Therefore the closed form solution is,
\begin{equation}
	\overline{F_{Y|X}}(y|{\bf x_k})=Q\biggl(\frac{\ln(\frac{1}{\gamma_m})-m_m}{\sigma_m}\biggr)
\end{equation}
where $m_m$ and $\sigma_m$ are the mean and standard deviation of resultant log normal RV derived in Section \ref{sec:cor_coe}.
Thus, (\ref{eqn:outage_macro}) can be rewritten as follows;
\begin{equation}\label{eqn:outage_macro_final}
	q_{m,{\bf r}}(\lambda_f)=\sum_{{\bf x_k}\in\mathcal{X}}Q\biggl(\frac{-\ln(\gamma_m)-m_m}{\sigma_m}\biggr)\mathbb{P}({\bf X}={\bf x}_k)
\end{equation}

\subsection{Femtocell Outage Probability}

The worst case scenario is when the FUE is at the edge of 
the 0-th femtocell and needs to be served by the 0-th FBS. The SIR for FUE is given as;
\pagebreak
\begin{equation}\label{eqn:sir_femto}
	\text{SIR}_{f,\vect{r}}=\frac{P_0\phi_0R_f^{-\beta}}{I_{f,m}+I_{f,f}}
\end{equation}

This expression contains a log normal RV $\phi_0$ instead of a Rayleigh RV in (\ref{eqn:sir_macro}). Hence its CCDF contains a weighted sum of ratio of log normal RV to a log normal RV and ratio of Rayleigh RV to log normal RV.
Continuing the previous steps, we obtain the outage probability for FUE as; 
\begin{equation}\label{eqn:outage_femto_final}
	q_{f,{\bf r}}(\lambda_f)=\sum_{{\bf x_k}\in\mathcal{X}}Q\biggl(\frac{-\ln(\gamma_f)-m_f}{\sigma_f}\biggr)\mathbb{P}({\bf X}={\bf x}_k)
\end{equation}
where $\gamma_f$ is the target SIR for any FUE. The calculation of $m_f$ and $\sigma_f$ is provided in section 
\ref{sec:cor_coe}.

\subsection{Total Transmission Capacity}

The total transmission capacity (TC) can be obtained using spatial throughput (ST), concurrent thriving transmissions per unit area, constrained with quality of service (QoS) requirement \cite{pap:spatialthroughput},\cite{pap:perfanatwotier}.
For a given $\lambda_f$, ST $\tau(\lambda_f)$ can be determined by the product of average successful probability and the transmission density such that,
 \begin{equation}\label{eqn:ST}
	\tau(\lambda_f)=\frac{1}{|H|}[1- q_m(\lambda_f)]+\lambda_f[1-q_f(\lambda_f)]
\end{equation}
where $q_m(\lambda_f)$ and $q_f(\lambda_f)$ are the average outage probabilities for MUE and FUE, respectively. Let us define the QoS constraint such that the allowable failure fraction for macrocell transmissions is $\epsilon_m$ and for femtocell transmissions $\epsilon_f$. Along with it, the expression (\ref{eqn:ST}) is modified and provides the TC as follows;
 \begin{equation}\label{eqn:TC}
	C=\frac{1}{|H|}[1- q_m(\overline{\lambda_f})]+\lambda_f[1-q_f(\overline{\lambda_f})]
\end{equation}
where $\overline{\lambda_f}=\min \bigl( q_m^{-1}(\epsilon_m), q_f^{-1}(\epsilon_f) \bigr)$ is the optimal femtocell density under the QoS requirement \cite{pap:perfanatwotier}.

\section{Approximating to Log Normal Distribution}\label{sec:combined}

\begin{figure}[!t]
	\centering
	\includegraphics[scale=0.26]{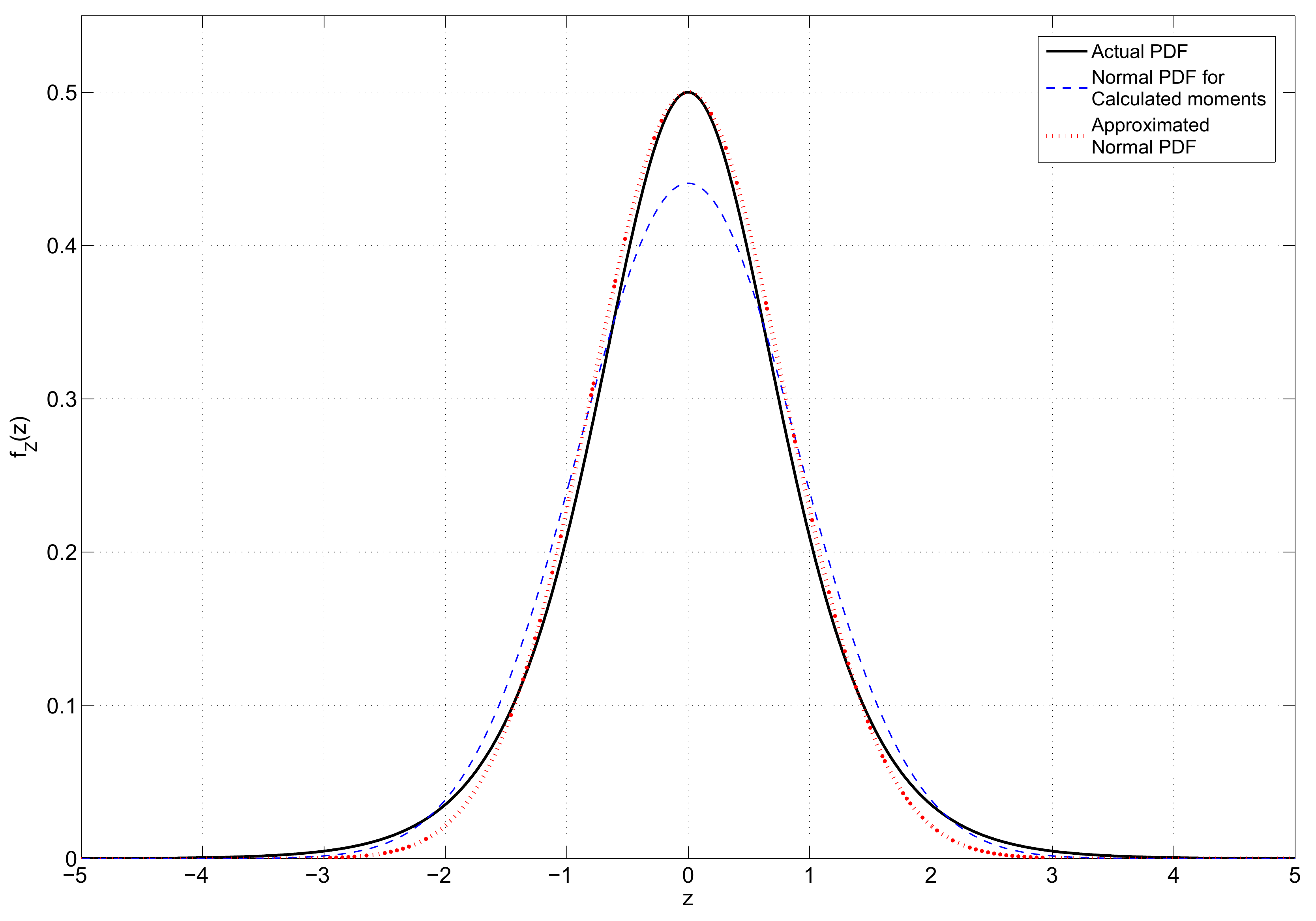}
	\caption{PDF: Ratio of two Rayleigh RVs}
	\label{fig:ral_ral_pdf}
\end{figure}

Deriving PDFs of ratios in Rayleigh and log normal faded environment are provided in \cite{pap:ralapprx},\cite{pap:logralapprx}.
Ref. \cite{pap:ralapprx} provides an approximation in Rayleigh only environment while \cite{pap:logralapprx} considers a system which has both Rayleigh and log normal fading with no difference in the indoor and outdoor links.
In our case, we approximate the ratios to log normal RVs and caluculate the numerical values.

\subsection{PDFs of ratios}\label{sec:apprxratioz}
We follow three main steps to obtain the approximated PDF. Initially we define a RV $Z$ where $e^Z$ is equal to the ratio. With this transformation we find the expression for $f_Z(z)$, i.e. PDF of $Z$. Then we calculate the mean $\mathbb{E}[Z]$ and the variance $\mathbb{V}[Z]$ from the actual PDF. We define the calculated PDF as $N(\mathbb{E}[Z],\mathbb{V}[Z])$; a normal PDF which is compared with the actual PDF. Finally, starting from the calculated PDF, we change the mean and the variance of normal distribution and calculate the point-wise difference from the actual PDF (point-wise errors). We select the normal distribution which has the minimum sum of errors as the approximated PDF.

\subsubsection{Ratio of Rayleigh RV to Rayleigh RV}\label{R_R}

Let $\psi_0$ and $\psi$ be i.i.d. Rayleigh RVs with parameter $\sigma$. We define the RV $Z$ s.t. $e^Z=\frac{\psi}{\psi_0}$.
With the conditional PDF $f_{Z|\psi_0}(z|\psi_0)=f_{\psi}(\psi)\frac{d\psi}{dz}$,
the PDF of $Z$ can be found using
\begin{equation}\label{eqn:pdfofral_ral}
	f_{Z}(z)={\int_0^\infty}f_{Z|\psi_0}(z|\psi_0)f_{\psi_0}(\psi_0)d\psi_0=\frac{2e^{2z}}{(1+e^{2z})^2}
\end{equation}

The mean and the variance of $Z$ are calculated using $f_{Z}(z)$, which are 0 and 0.6179 respectively. Hence the normal PDF for calculated moments is $N(0,0.6197)$. The actual PDF and calculated PDF are plotted in Fig. \ref{fig:ral_ral_pdf} along with the approximated PDF, which is obtained by iterations. Approximated PDF is $N(0,0.7979^2)$, thus the RV $\frac{\psi}{\psi_0}$ can be approximated to a $LN(0,0.7979^2)$.

\subsubsection{Ratio of Rayleigh RV to Log Normal RV}\label{L_R}

Let $\psi$ be a Rayleigh RV with parameter $\sigma$ and $\phi$ be a log normal RV with parameters 0 and $\sigma$. RV $Z$ is defined as $e^Z=\frac{\phi}{\psi}$. With a similar approach we can obtain the PDF of $Z$ which is,
\begin{align}
	\nonumber &f_{Z}(z)=\frac{1}{\sqrt{2\pi}\sigma^2}\exp\{-2(z-\sigma^2)\}\times\\
	&~~~~~~~~{\int_{-\infty}^\infty}\exp\big\{\frac{-\mu^2-e^{-2(z-2\sigma^2)}e^{2\mu}}{2\sigma^2}\big\}d\mu
\end{align}
where $\mu=\ln\psi+(z-2\sigma^2)$.
This integration is evaluated using Trapezoidal rule with the simplification of $\sigma=1$ (\ref{eqn:trapz}).
\begin{align} \label{eqn:trapz}
	f_{Z}(z)&=\frac{e^{-2z}}{\sqrt{2\pi}}{\int_{-\infty}^\infty}\exp\big\{\frac{-\mu^2-e^{-2(z-2)}e^{2\mu}}{2}\big\}d\mu
\end{align}
$\mathbb{E}~[Z]=-0.058$ and $\mathbb{V}[Z]=1.4112^2$ are calculated from $f_{Z}(z)$.
The normal PDF for above calculated moments is $N(-0.058,1.4112^2)$ and the obtained approximated normal PDF is $N(-0.143,1.1673^2)$. Therefore the ratio is approximated into $LN(-0.143,1.1673^2)$.

For the ratio of Rayleigh RV to log normal RV, which is now $e^{Z_{new}}=\frac{\psi}{\phi}=e^{-Z}$, we can deduce that it can be approximated to $LN(0.143,1.1673^2)$.

\subsection{Weighted summation}\label{sec:cor_coe}

In (\ref{eqn:ccdf_condi}), we need to calculate PDF of the weighted sum of RVs. For the ratio of log normal RVs, this can be accomplished using Fenton - Wilkinson's method \cite{pap:perfanatwotier}. 
We define a RV $Z_k$ s.t. $e^{Z_k}$ is equal to one of $\frac{\psi_k}{\psi_0}$, $\frac{\phi_k}{\psi_0}$ (when considering MUE) or $\frac{\psi_k}{\phi_0}$, $\frac{\phi_k}{\phi_0}$ (when considering FUE) where $\psi$ and $\phi$ are Rayleigh and log normal RVs, respectively. Then the required weighted sum can be rewritten in the form of $e^{Z_x}=\sum_{k{\in}\Omega\cup\Lambda}\xi_ke^{Z_k}$ where $\xi_k$ are the weights.
The resultant log normal RV is $e^{Z_x}\sim LN(m_x,\sigma^2_x)$ where the parameters are given as below.
\begin{align}\label{eqn:resul_moment_calc}
	\nonumber \mathbb{E}~[e^{Z_m}]=&\sum_{k\in\Omega\cup\Lambda}\xi_ke^{\delta_{kk}\sigma_{Z_k}^2}\\
	\nonumber \mathbb{E}~[e^{2Z_m}]=&\sum_{k\in\Omega\cup\Lambda}\biggl(\xi_k^2e^{2\sigma_{Z_k}^2}+\\
	&~~~~~~~\sum_{\substack{l\in\Omega\cup\Lambda\\l{\neq}k}}\xi_k\xi_le^{\frac{\sigma_{Z_k}^2+\sigma_{Z_l}^2+2\delta_{kl}\sigma_{Z_k}\sigma_{Z_l}}{2}}\biggr)
\end{align}
\begin{align}\label{eqn:resul_moment}
	\nonumber m_x&=2\ln(\mathbb{E}~[e^{Z_m}])-0.5\ln(\mathbb{E}~[e^{2Z_m}])\\
	\sigma_x^2&=\ln(\mathbb{E}~[e^{2Z_m}])-2\ln(\mathbb{E}~[e^{Z_m}])
\end{align}

For $Z_1$  and $Z_2$, $\delta_{12} = \frac{\mathbb{E}~[(Z_1-\overline Z_1)(Z_2-\overline Z_2)]}{\sigma_{Z_1}\sigma_{Z_2}}$ is the correlation coefficient. Using normal PDFs with calculated moments or approximated normal PDFs obtained in section \ref{R_R},\ref{L_R}, two possible values can be obtained for $\delta$ which we define as $\delta_{cal}$ and $\delta_{apprx}$, respectively.

\begin{figure}[!t]
	\centering
	\includegraphics[scale=0.26]{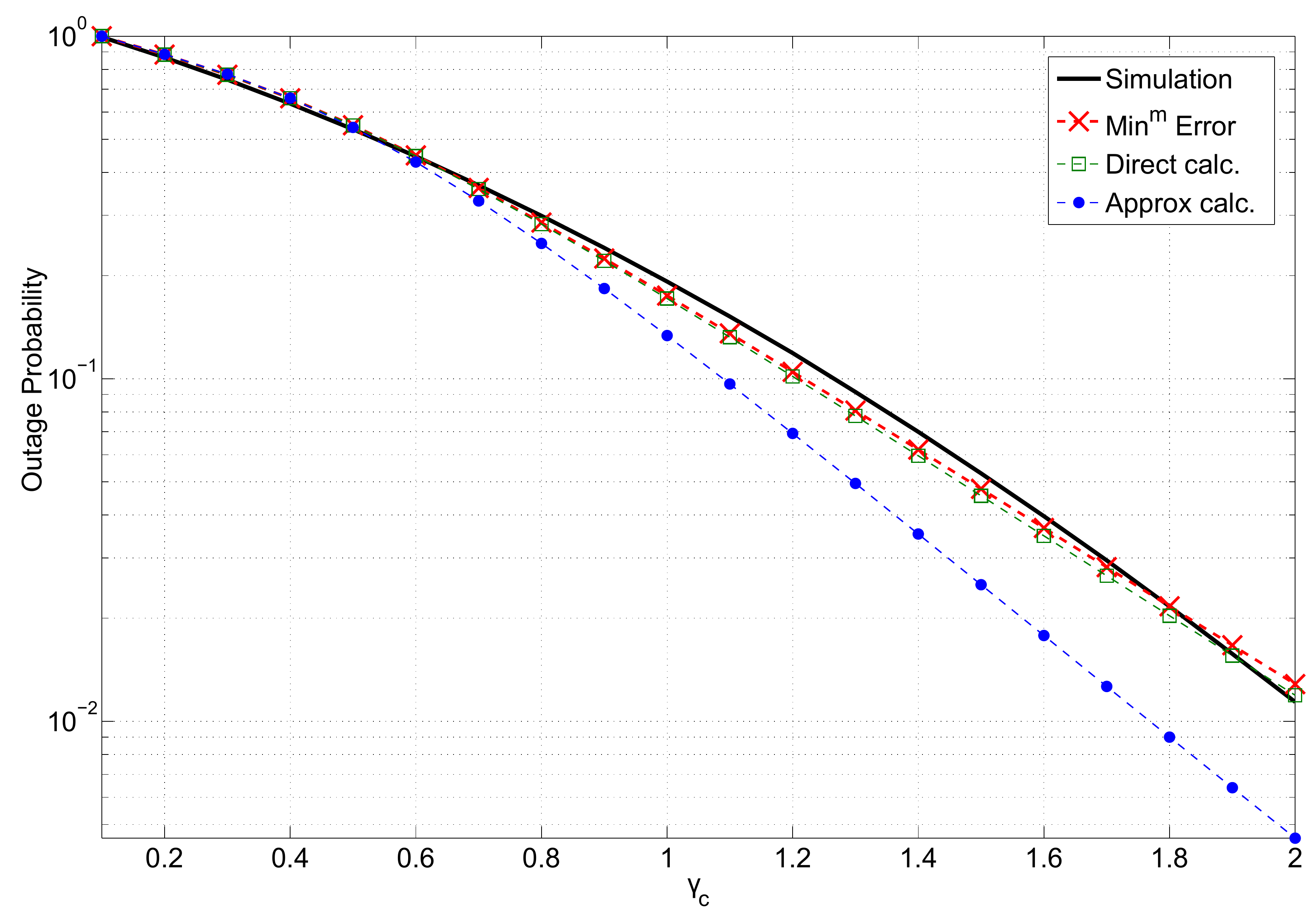}
	\caption{Comparison on weighted sum of approximated RVs of $\frac{\psi_1}{\psi_0}$ and $\frac{\psi_2}{\psi_0}$, $\delta_{cal}=0.5017$, $\delta_{apprx}=0.6495$ and $\delta_{minerr}=0.4857$}
	\label{fig:R_R_mod_simu}
\end{figure}

Then we perform a simulation for the weighted summation $Z$ and plot Pr$(Z>\gamma)$ versus $\gamma$. It is compared with the analytical values which are obtained for both correlation coefficients, $\delta_{cal}$ and $\delta_{apprx}$. Finally, using a number of iterations we obtain the $\delta=\delta_{minerr}$, which provides the minimum square error with the simulated curve. A comparison is shown in Fig. \ref{fig:R_R_mod_simu}.
The obtained $\delta_{minerr}$ values from calculations and simulations for the required combinations are given in Table \ref{tab:delta}.

\begin{table}[ht]
\caption{correlation coefficients ($\delta_{minerr}$)}
\centering
\begin{center}
\begin{tabular}{|c|c|c||c|c|c|}
\hline
 & $\frac{\psi_1}{\psi_0}$ & $\frac{\phi_1}{\psi_0}$ &  & $\frac{\psi_1}{\phi_0}$ & $\frac{\phi_1}{\phi_0}$ \\
\hline
$\frac{\psi_2}{\psi_0}$ & 0.4857 & 0.3879 & $\frac{\psi_2}{\phi_0}$ & 0.5252 & 0.4856 \\
\hline
$\frac{\phi_2}{\psi_0}$ & 0.3879 & 0.4895 & $\frac{\phi_2}{\phi_0}$ & 0.4856 & 0.5\\
\hline
\end{tabular}
\end{center}
\label{tab:delta}
\end{table}

\section{Numerical Results}\label{sec:results}

With the following parameters, analytical results are compared with simulations. We select $R_m=500m$ and $R_f=20m$.
The central macrocell has an average of 20 femtocells with the intensity of $\lambda_f$ and is surrounded by two rings of interfering macrocells. All BSs have omni directional antennas and, transmission powers for MBSs are $P_m=P_j=50$ dBm ($j\in\Omega_m$) and for FBSs $P_f=P_i=22,25$ dBm ($i\in\Lambda$). For the path loss calculation, $\alpha=4,~\beta=3$ and $W=12$ dB. Also we consider $\gamma_m,\gamma_f$ to be 1 and 10, respectively.

\begin{figure}[!t]
	\centering
	\includegraphics[scale=0.26]{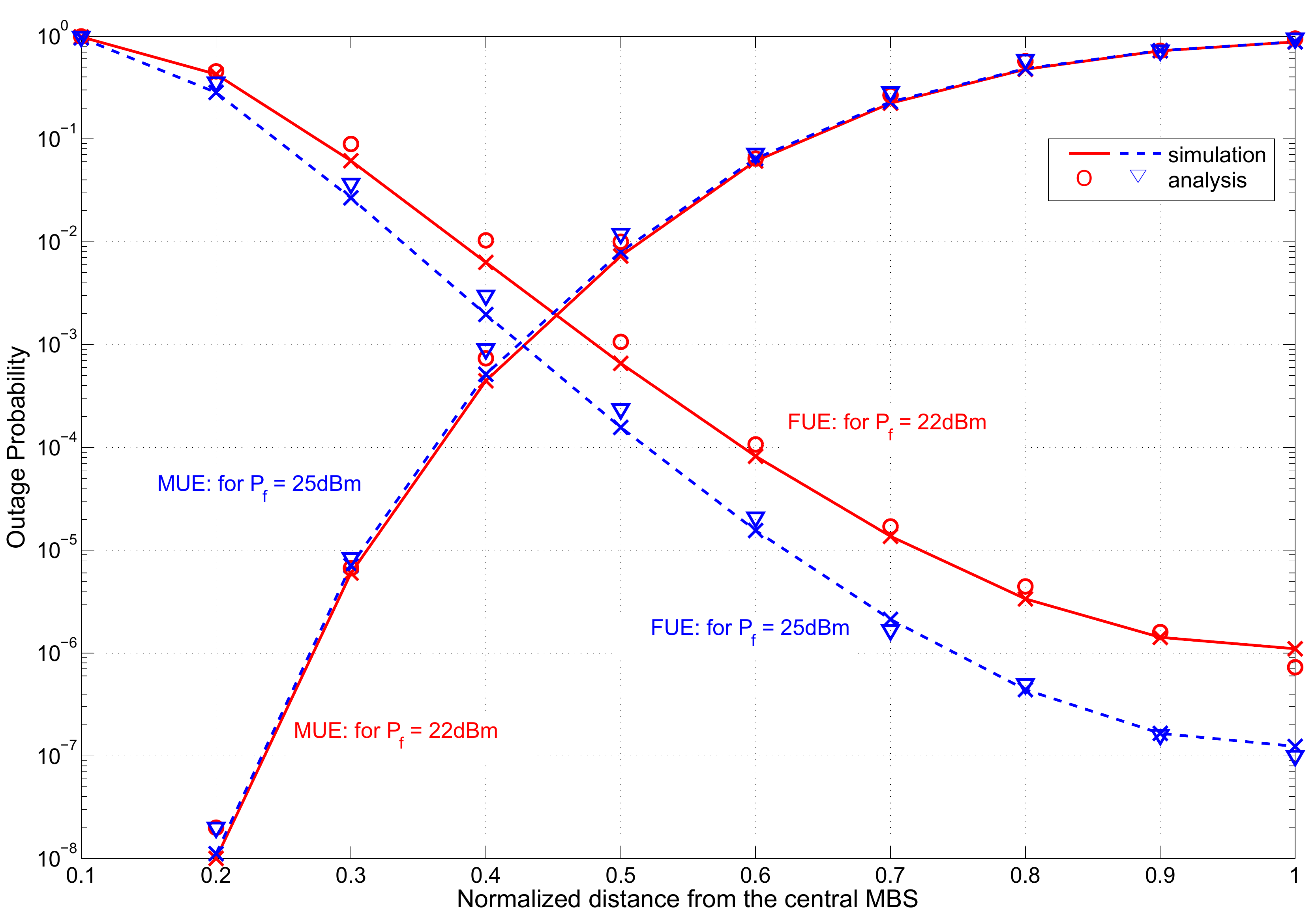}
	\caption{Outage probability; $P_f=22,25$dBm, $P_m=50$dBm \& $R_m=500$m.}
	\label{fig:fBSchange}
\end{figure}

First we analyze the outage probability (OP) for two different transmission power levels of FBSs as illustrated in Fig. \ref{fig:fBSchange}. With the distance the received power for MUE degrades and therefore its OP increases. When FBS power is increased, the amount of interference to MUE becomes higher along with the OP.
For FUE, as the distance from the central BS increases, the interference from 0-th MBS reduces. Thus FUE OP decreases with the distance. When we increase the FBS power, though it effects both the signal and the interference of FUE, the interference attenuates more due to wall penetration.
Hence we can observe a lower FUE outage. 

\begin{figure}[!t]
	\centering
	\includegraphics[scale=0.26]{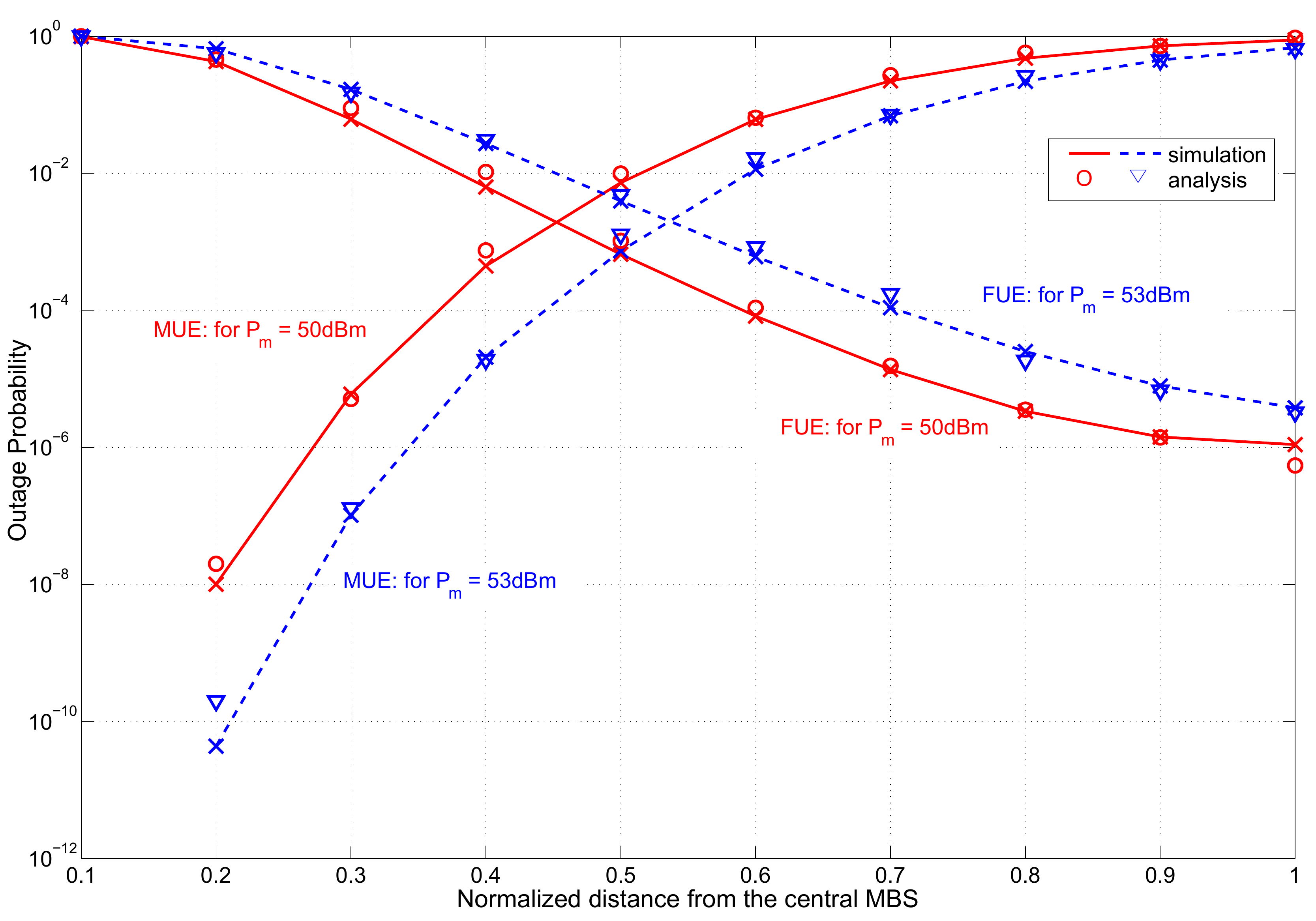}
	\caption{Outage probability; $P_f=22$dBm, $P_m=50,53$dBm \& $R_m=500$m.}
	\label{fig:mBSchange}
\end{figure}

Fig. \ref{fig:mBSchange} shows the behavior for different MBS transmission powers. Though it increases both signal and interference for MUE, due to the inverse relation of the distance, the effect from signal power on SIR is larger than the effects from the interference. Consequently the MUE OP decreases. However, this gain of MBS gives additional interference to the FUE. Thus we can see the rise of the FUE outage probability.

\begin{figure}[!t]
	\centering
	\includegraphics[scale=0.26]{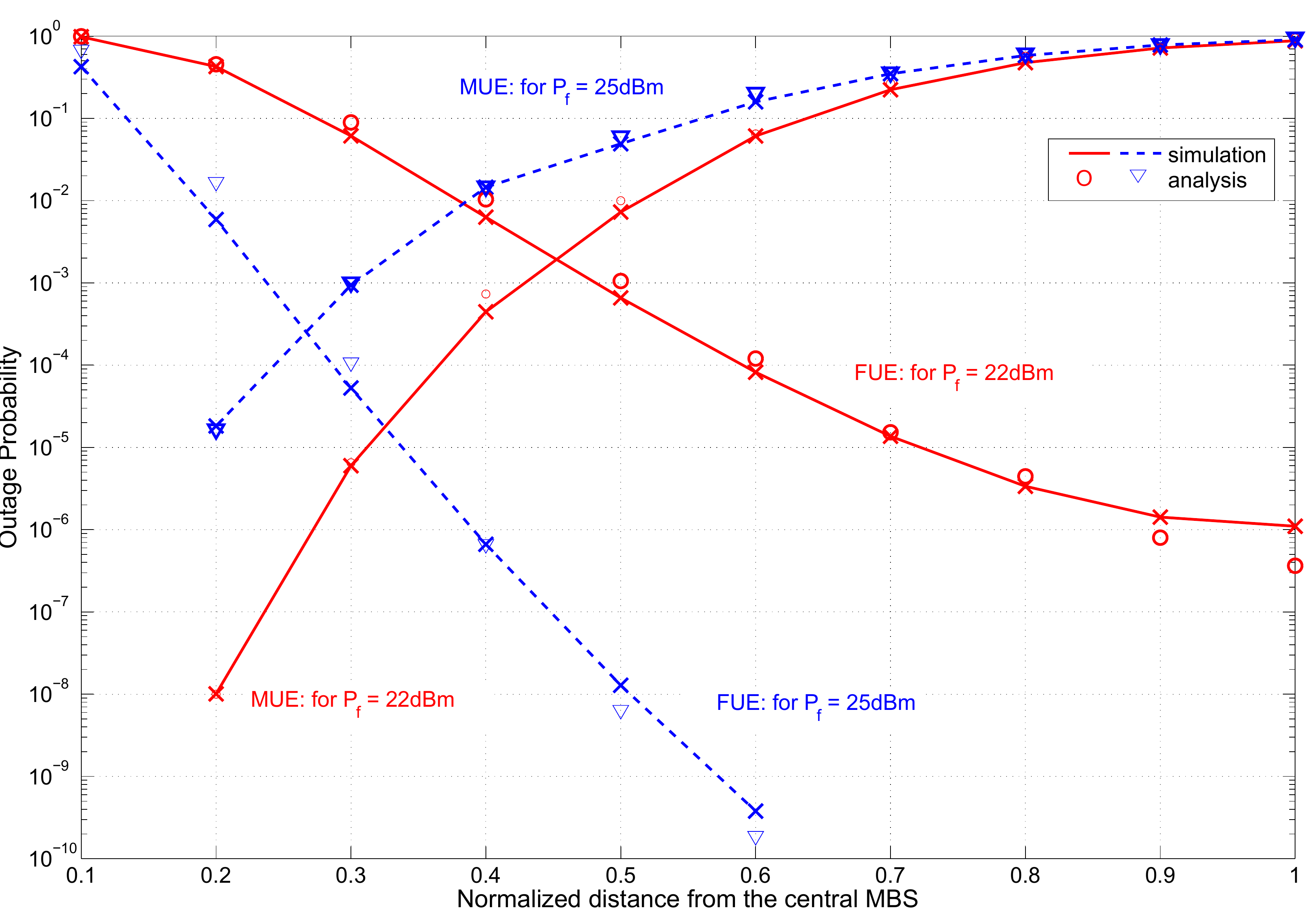}
	\caption{Outage probability: $P_f=22$dBm, $P_m=50$dBm \& $R_m=0.5,1$km.}
	\label{fig:Rcchange}
\end{figure}

Fig \ref{fig:Rcchange} gives the OP variation for two macrocell sizes. Both the signal strength for MUE and the interference from MBS to FUE are low if they are away from the MBS when the cell is large. Therefore, the OP is higher for the MUE and smaller for FUEs.
 
\begin{figure}[!t]
	\centering
	\includegraphics[scale=0.26]{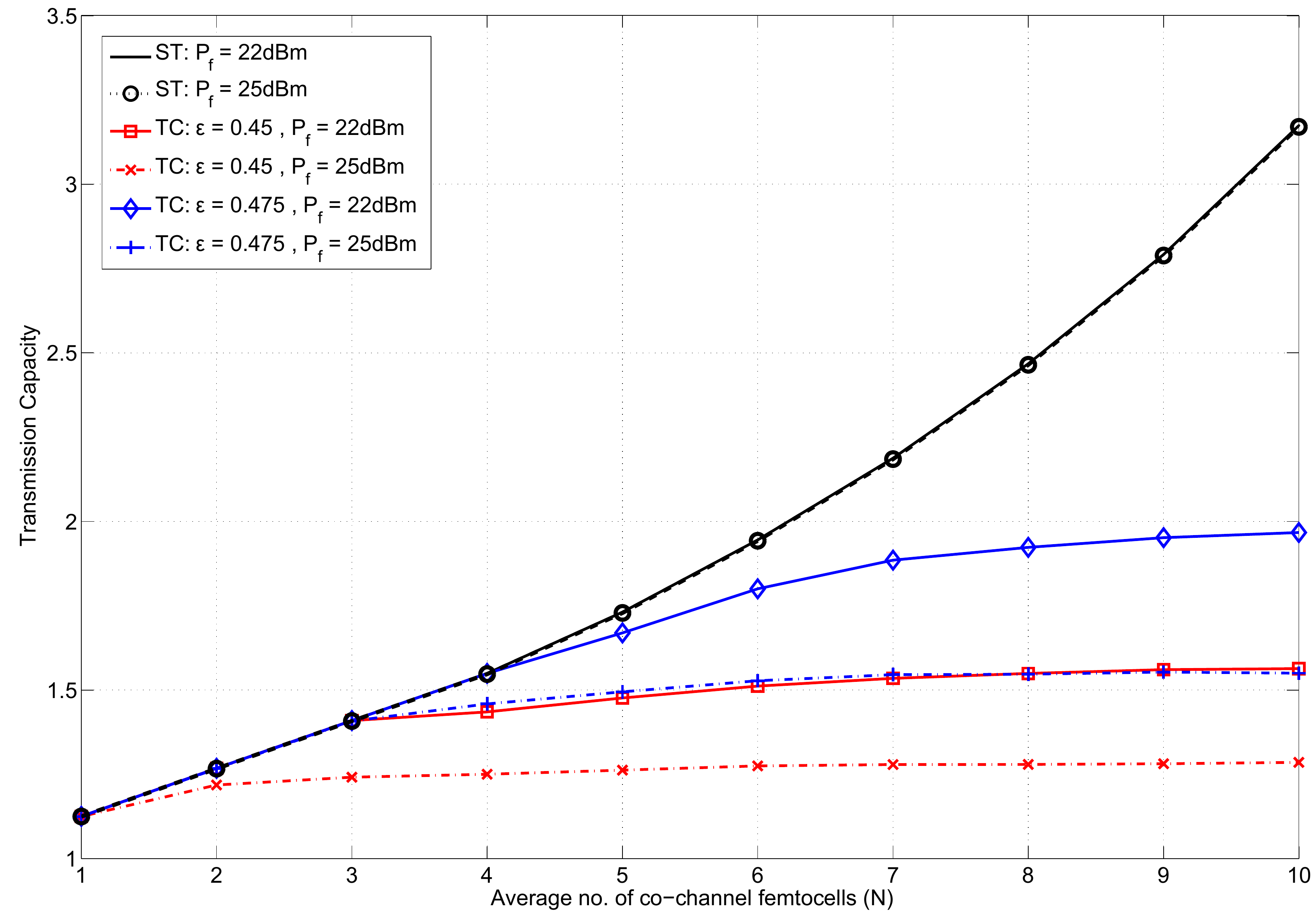}
	\caption{Transmission capacity (TC) and spatial throughput (ST) comparison}
	\label{fig:capacity}
\end{figure}

For the capacity calculation, we set $\epsilon_f=0.045$ and consider two scenarios for MUE, $\epsilon_m=0.45$ and $\epsilon_m=0.475$. Without the QoS requierment ($\overline{\lambda_f}$), we calculate the ST and then use (\ref{eqn:TC}) to find the TC with the constraints mentioned above. Fig. \ref{fig:capacity} illustrates the behavior of TC and ST versus the average number of co-channel femtocells.  With the QoS requirement the TC value becomes a constant. As the constraint is relaxed, i.e. higher failures are allowed, TC increases, which conveys the idea that TC has an inverse relation with QoS.
A higher transmission power in FAPs provides better service to FUE. However, it increases the cross-tier interference resulting an overall reduction in the system capacity. Thus we can see that a small number of co-channel femtocells per site would be suitable in order to maintain the TC at an optimal value.
\section{Conclusion}\label{sec:conclu}
In this paper, we provide derivations for per-tier outage probabilities and the capacity of a system with co-channel femtocell network overlaid with a macrocell. To model the system closer to realistic conditions, we use different propagation models for indoor and outdoor, Rayleigh and log normal shadowed, respectively. Using those, for analytical derivations, we present approximations for PDFs and weighted sums for the ratios of Rayleigh RVs and Rayleigh to log normal RVs. The simulated results are compared with analytical results and the system behavior is analyzed. The installation of co-channel femtocells is feasible, though limited by the capacity with QoS constraints. Therefore maintaining an optimal number of co-channel femtocells with aid of using multiple sub carriers and cell partitioning  will effective in terms of both cost and the quality.

\bibliographystyle{IEEEtran}
\bibliography{IEEEabrv,paper}

\end{document}